\documentclass{PoS}

\title{Measurement of $\hat{q}$ in RHI collisions using di-hadron correlations}

\ShortTitle{Measurement of $\hat{q}$ in RHI collisions}

\author{\speaker{M. J. Tannenbaum}\thanks{Supported by U.~S. Department of Energy, Contract No. DE-SC0012704.}\\
Physics Dept., 510c, Brookhaven National Laboratory, Upton, NY 11973-5000,USA\\E-mail: \email{mjt@bnl.gov}}

\abstract{In the BDMPSZ model, the energy loss of an outgoing parton in a medium $-dE/dx$ is the transport coefficient $\hat{q}$ times $L$ the length traveled. This results in jet quenching, which is well established. However BDMPSZ also predicts an azimuthal broadening of di-jets also proportional to $\hat{q}L$ which has so far not been observed.
The azimuthal width of the di-hadron correlations in p$+$p collisions, beyond the fragmentation transverse momentum, $j_T$, is dominated by $k_T$, the so-called intrinsic transverse momentum of a parton in a nucleon, which can be measured. The broadening should produce a larger $k_T$ in A$+$A than in p$+$p collisions. This presentation introduces the observation that the $k_T$ measured in p$+$p collisions for di-hadrons with $p_{Tt}$ and $p_{Ta}$ must be reduced to compensate for the energy loss of both the trigger and away parent partons when comparing to the $k_T$ measured with the same di-hadron $p_{Tt}$ and $p_{Ta}$ in A$+$A  collisions. This idea is applied to a recent STAR di-hadron measurement in Au$+$Au at \sqsn=200 GeV, \Journal{\PLB}{760}{689}{2016}, with result $<{\hat{q}L}>=2.1\pm 0.6$ GeV$^2$. This is more precise but in agreement with a theoretical calculation of $<{\hat{q}L}>=14^{+42}_{-14}$ GeV$^2$ using the same data. Assuming a length $<{L}>\approx 7$ fm for central Au$+$Au collisions the present result gives $\hat{q}\approx 0.30\pm 0.09$ GeV$^2$/fm, in fair agreement with the JET collaboration result from single hadron suppression of $\hat{q}\approx 1.2\pm 0.3$ GeV$^2$/fm at an initial time \mbox{$\tau_0=0.6$ fm/c} in Au$+$Au collisions at $\sqrt{s_{NN}}=200$ GeV.
There are several interesting details to be discussed: for a given $p_{Tt}$ the $<{\hat{q}L}>$ seems to decrease then vanish with increasing $p_{Ta}$; the di-jet spends a much longer time in the medium ($\approx 7$ fm/c) then $\tau_0=0.6$ fm/c which likely affects the value of $\hat{q}$ that would be observed.
}

\FullConference{
12th International Workshop on High-pT Physics in the RHIC/LHC Era\\
		2-5 October, 2017\\
		University of Bergen, Bergen, Norway}

\usepackage{graphicx}	
\usepackage{xspace}     
\usepackage{amsmath}
\usepackage {bm}  

\newcommand{\sqs}{\mbox{$\sqrt{s}$}\xspace}
\newcommand{\sqsn}{\mbox{$\sqrt{s_{_{NN}}}$}\xspace}

\def\lsim{\raise0.3ex\hbox{$<$\kern-0.75em\raise-1.1ex\hbox{$\sim$}}}
\def\gsim{\raise0.3ex\hbox{$>$\kern-0.75em\raise-1.1ex\hbox{$\sim$}}}

\def\mean#1{\left<#1\right>}
\def\Journal#1#2#3#4{ {\it{#1}} {\bf #2}, #3 (#4)}

\def\EPJC{{Eur. Phys. J.}\ {\rm C}}

\def\JPG{{J. Phys.}\ {\rm G}}

\def\NPA{{Nucl. Phys.}\ {\rm A}}
\def\NPB{{Nucl. Phys.}\ {\rm B}}
\def\PLB{{Phys. Lett.}\ {\rm B}}
\def\PLC{Phys. Repts.\ }
 
\def\PRL{Phys. Rev. Lett.\ }
\def\PRD{{Phys. Rev.}\ {\rm D}}
\def\PRC{{Phys. Rev.}\ {\rm C}}
\def\ARNPS{{Ann. Rev. Nucl. Part. Sci.\ }}

\def\la{\left< }
\def\ra{\right> }
\def\jt#1{\ensuremath{j_{T\rm #1}}}
\def\meankv#1{\ensuremath{\la#1^2\ra}}
\def\rms#1{\meankv{#1}}

\def\QGP{\Red Q\Blue G\Green P\Black}
\def\QCD{\Red Q\Green C\Blue D\Black}

\begin{document}
\section{Jet Quenching: the first \QCD\ based prediction BDMPSZ~\cite{BSZ2000}} 
\begin{figure}[!h]  
\begin{center}
\raisebox{+2.0pc}{{\footnotesize a)}\hspace*{-0.05pc}\includegraphics[width=0.42\textwidth]{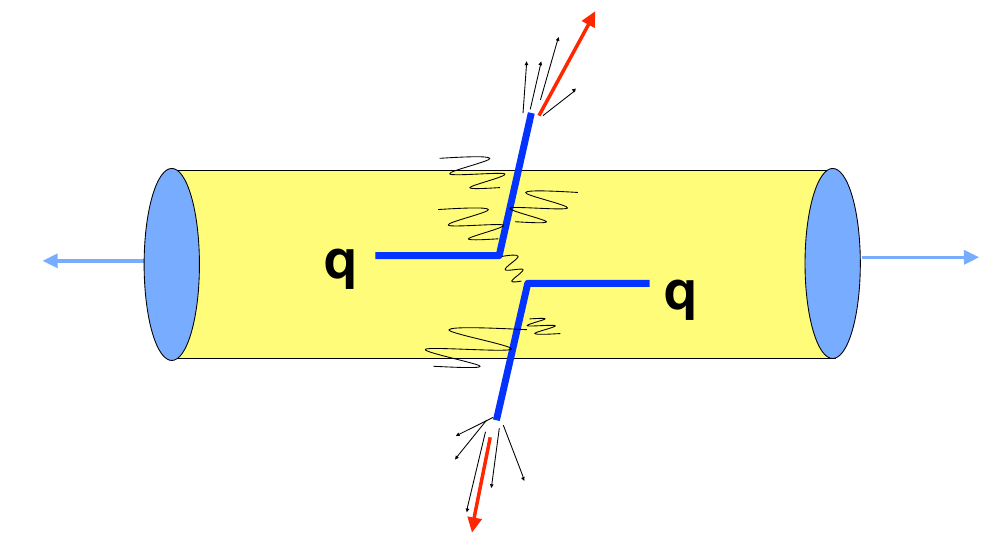}}
\raisebox{-0.0pc}{{\footnotesize b)}\hspace*{-0.2pc}\includegraphics[width=0.54\textwidth]{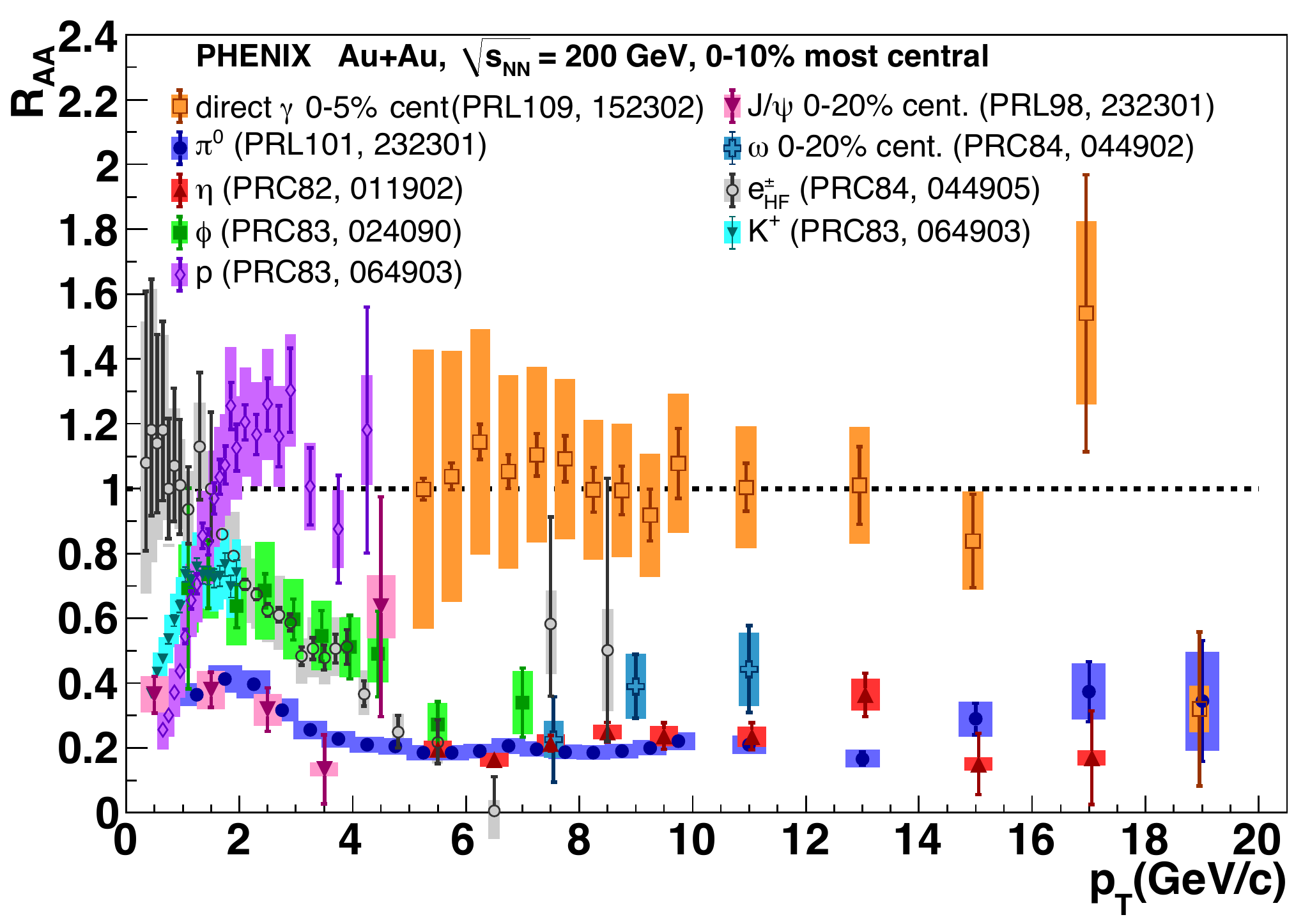}}
\caption[]{\footnotesize a) Schematic of $q+q$ scattering with scattered quarks losing energy in the medium. b) Suppression,  $R_{AA}(p_T)$, for all identified particles so far measured by PHENIX in Au$+$Au central collisions at $\sqsn=200$ GeV.}   
\label{fig:suppression}
\end{center}
\end{figure}

The first prediction of how to detect the \QGP\ was via $J/\Psi$ suppression~\cite{MatsuiSatz} in 1986. However the first \QCD\ based prediction for detecting the \QGP\ was BDMPSZ Jet Quenching~\cite{BSZ2000}.  This is produced by the energy loss, via LPM coherent radiation of gluons, of an outgoing parton with color charge fully exposed in a medium with a large density of similarly exposed color charges (i.e. the \QGP) (Fig.~\ref{fig:suppression}a). Jet quenching was observed quite early at RHIC by suppression of high $p_T$ $\pi^0$~\cite{PXPRL88},  with lots of subsequent evidence (Fig.~\ref{fig:suppression}b). It is interesting to note that all identified hadrons generally have different $R_{AA}$ for $p_T\leq5$ GeV/c but tend to converge to the same value for $p_T\gsim5$ GeV/c. The fact that direct-$\gamma$ are not suppressed indicates that suppression is a medium effect on outgoing color-charged partons as predicted by BDMPSZ~\cite{BSZ2000}.  
\subsubsection{But the BDMPSZ model has two predictions}
\begin{figure}[!h]  
\begin{center}
\raisebox{+0.0pc}{\includegraphics[width=0.60\textwidth]{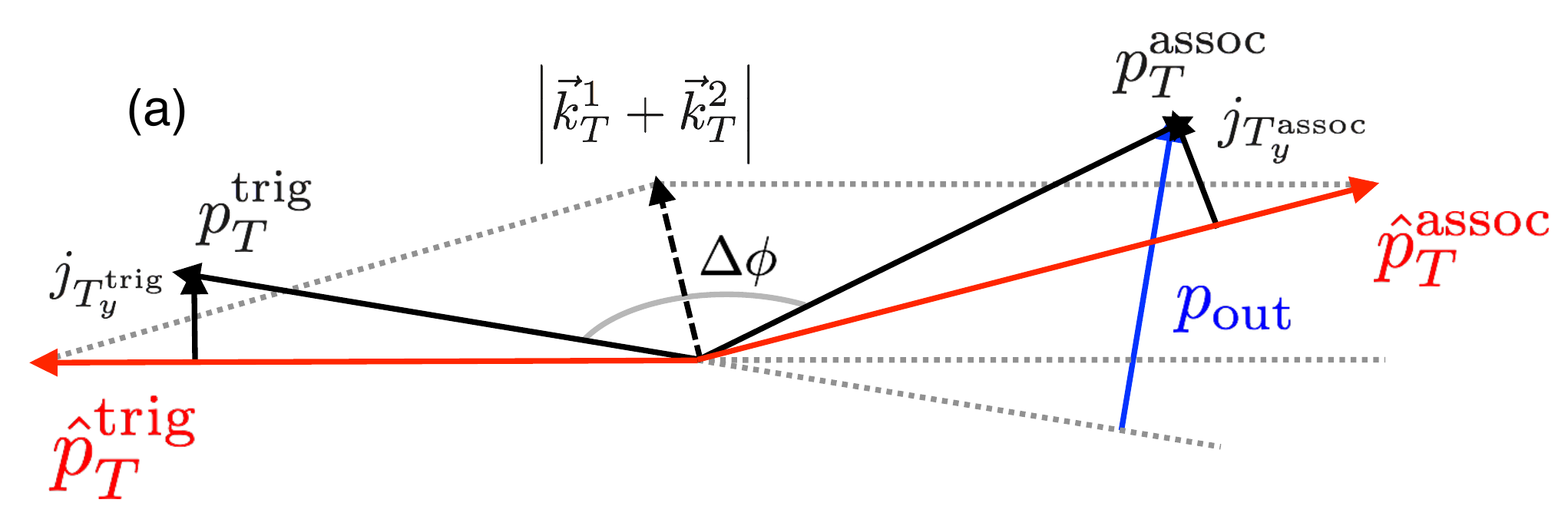}}\hspace*{0.2pc}
\caption[]{\footnotesize Initial configuration: trigger jet $\hat{p}_{Tt}$, associated (away) jet $\hat{p}_{Ta}$  with $k_T$ effect (dashed arrow) and fragments $p_{Tt}$ and $p_{Ta}$, with fragmentation transverse momentum $j_{T_y}$, and $p_{out}=p_{Ta}\sin(\pi-\Delta\phi)$.   }   
\label{fig:kTprime}
\end{center}
\end{figure}

\noindent(I) The energy loss of the outgoing parton, $-dE/dx$,  
per unit length ($x$) of a medium with total length $L$, is proportional to the total 4-momentum transfer-squared, $q^2(L)$, with the form:\vspace*{-0.5pc}
\begin{equation}\frac{-dE} {dx}\simeq \alpha_s \langle{q^2(L)}\rangle=\alpha_s\, \mu^2\, L/\lambda_{\rm mfp} 
=\alpha_s\, \hat{q}\, L\qquad \label{eq:dEdx} \end{equation}
where $\mu$, is the mean momentum transfer per collision, and the transport coefficient 
{$\hat{q}=\mu^2/\lambda_{\rm mfp}$} is the 4-momentum-transfer-squared to the medium per mean free path, $\lambda_{\rm mfp}$.\\[0.5pc] 
\noindent(II) Additionally, the accumulated momentum-squared, $\mean{p^2_{\perp W}}$ transverse to a parton traversing a length $L$ in the medium  is well approximated by\vspace*{-0.5pc} 
\begin{equation}\mean{p^2_{\perp W}}\approx\langle{q^2(L)}\rangle=\hat{q}\, L \qquad \mbox{\rm so that}\qquad  \mean{\hat{q} L}/2=\mean{k_{T}^2}_{AA}-\mean{k{'}_{T}^2}_{pp} \label{eq:broadening}\end{equation} since only the component of $\mean{p^2_{\perp W}}$ $\perp$ to the scattering plane affects $k_T$. This is called azimuthal broadening. Here (see Fig.~\ref{fig:kTprime}) $k_T$ denotes the intrinsic transverse momentum of a parton in a proton plus any medium effect and $k{'}_T$ denotes the reduced value correcting for the lost energy of the scattered partons in the \QGP, a new idea this year~\cite{MJTPLB771}.

\begin{figure}[!h]  
\begin{center}
\raisebox{+0.0pc}{{\footnotesize a)}\hspace*{-0.05pc}\includegraphics[width=0.43\textwidth]{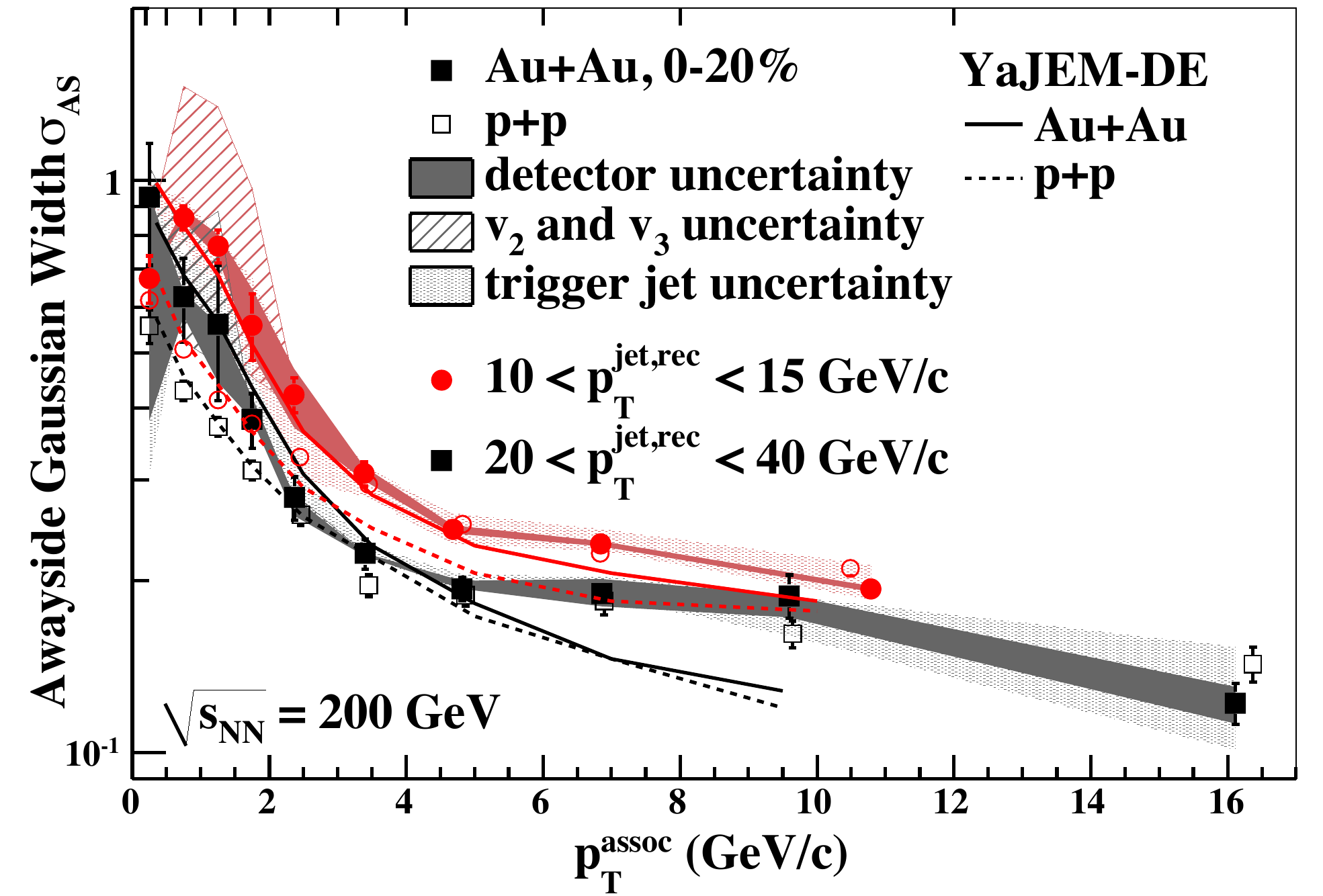}}
\raisebox{-0.0pc}{{\footnotesize b)}\hspace*{-0.05pc}\includegraphics[width=0.53\textwidth]{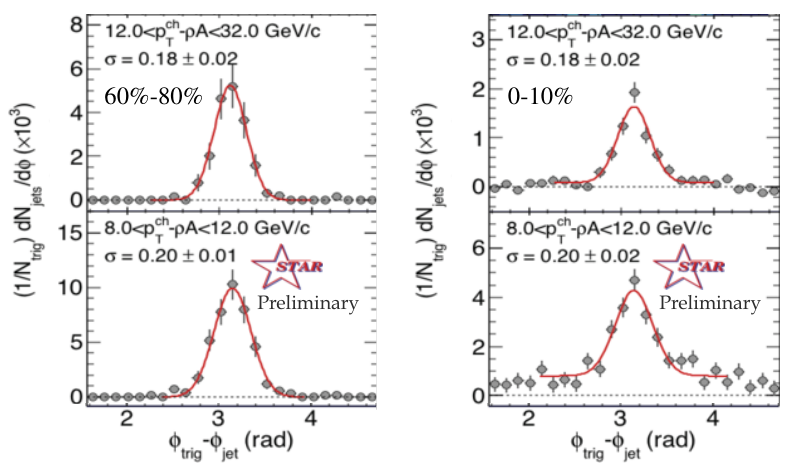}}
\caption[]{\footnotesize a) STAR measurement of the Gaussian widths $\sigma_{\rm AS}$ of away-side
hadron peaks triggered by a jet in collisions of Au$+$Au (solid symbols) and p$+$p (open symbols) at \sqsn=200 GeV~\cite{STARJhPRL112}. b) Away-peaks in STAR di-jet measurement for two $\hat{p}_{Tt}$ ranges in Au$+$Au at \sqsn=200 GeV: (left) peripheral, (right) central collisions, with the same $\sigma$~\cite{JacobsNPA956}.}   
\label{fig:notbroad}
\end{center}
\end{figure}

Even though jet quenching has been established and confirmed for more than 15 years, many experiments have tried to find azimuthal broadening at RHIC e.g. Fig.~\ref{fig:notbroad} \cite{STARJhPRL112}, \cite{JacobsNPA956}, but have not been able to observe the effect because of systematic uncertainties.

\subsection{ Understanding $k_T$ and $k{'}_T$.}
Following the methods of Feynman,Field and Fox~\cite{FFFNPB128}, CCOR~\cite{CCORjTkT} and PHENIX~\cite{ppg029}, the $\mean{k^2_T}$ for di-hadrons is computed from Fig.~\ref{fig:kTprime} as:
\begin{equation}\sqrt{\mean{k^2_T}}=\frac{\hat{x}_h}{\mean{z_t}}\sqrt{\frac{ \mean{p^2_{\rm out}}-(1+{x_h^2})\rms{j_T}/2}{x_h^2}} \label{eq:kTcalc}\end{equation}
where ${p}_{Tt}$, ${p}_{Ta}$ are the transverse momenta of the trigger and away particles, \mbox{$x_h=p_{Ta}/p_{Tt}$}, $\Delta\phi$ is the angle between ${p}_{Tt}$ and ${p}_{Ta}$ and  $p_{\rm out}\equiv p_{Ta} \sin(\pi-\Delta\phi)$.  The di-hadrons are assumed to be fragments of jets with transverse momenta $\hat{p}_{Tt}$ and $\hat{p}_{Ta}$ with ratio $\hat{x}_h=\hat{p}_{Ta}/\hat{p}_{Tt}$. ${z_t}\simeq p_{Tt}/\hat{p}_{Tt}$ is the fragmentation variable, the fraction of momentum of the trigger particle in the trigger jet. $j_T$ is the jet fragmentation transverse momentum and we have taken $\rms{\jt{ay}}\equiv\rms{\jt{a\phi}}=\rms{\jt{t\phi}}=\rms{j_T}/{2}$. The variable $x_h$ (which STAR calls $z_T$) is used as an approximation of the variable \mbox{$x_E=x_h\cos(\pi-\Delta\phi)$} from the original terminology at the CERN ISR where $k_T$ was discovered and measured 40 years ago.

\begin{figure}[!h]  
\begin{center}
\raisebox{+0.0pc}{\includegraphics[width=0.88\textwidth]{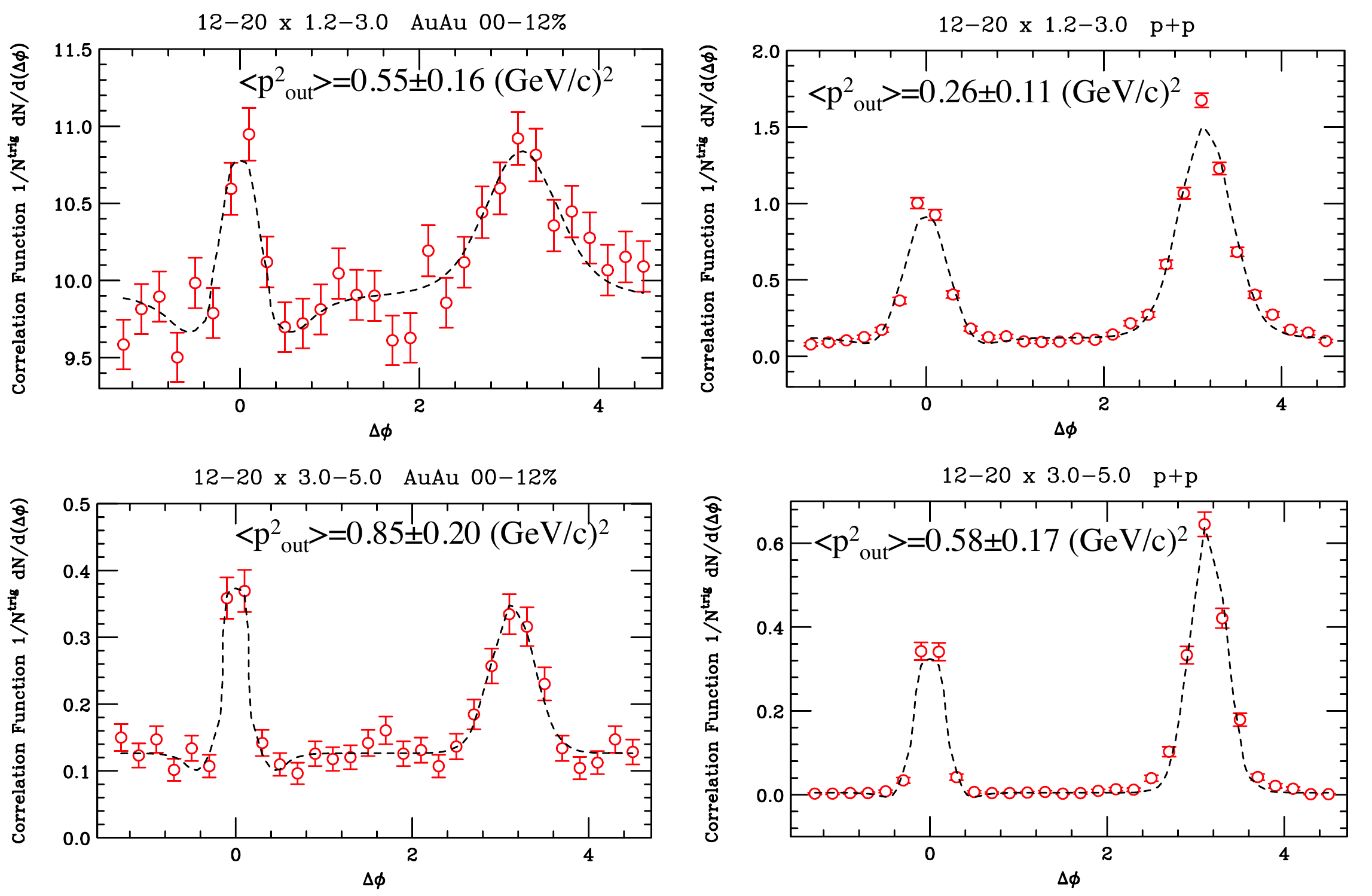}}
\caption[]{\footnotesize Fits~\cite{MJTPLB771} to STAR $\pi^0$-hadron correlation functions~\cite{STARPLB760}: Gaussian in $\Delta\phi$ on trigger side  ($\Delta\phi\approx0$), and Gaussian in $p_{\rm out}$ on away-side with fitted values of $\mean{p^2_{\rm out}}$ indicated. }   
\label{fig:poutfitstostar}
\end{center}
\end{figure}

A recent STAR paper~\cite{STARPLB760} on $\pi^0$-hadron correlations in $\sqsn=200$ GeV Au$+$Au 0-12\% central collisions had very nice correlation functions for large enough $12\leq p_{Tt}\leq 20$ GeV/c so that the $v_2$, $v_3$ modulation of the background was negligible (Fig.~\ref{fig:poutfitstostar}). I made fits to these data~\cite{MJTPLB771} to determine $\mean{p^2_{\rm out}}$ so that I could calculate $k_T$ in p$+$p and Au$+$Au using Eq.~\ref{eq:kTcalc}. The results for $3\leq p_{Ta}\leq 5.0$ GeV/c were 
$\sqrt{\mean{k^2_T}}=2.5\pm0.3$ GeV/c for p$+$p and $\sqrt{\mean{k^2_T}}=1.4\pm0.2$ GeV/c, for Au$+$Au, exactly the opposite of azimuthal broadening (Eq.~\ref{eq:broadening})!  
 
After considerable thought, I finally figured out what the problem was and introduced the new $k^{'}_T$~\cite{MJTPLB771}.  For a di-jet produced in a hard scattering, the initial $\hat{p}_{Tt}$ and $\hat{p}_{Ta}$ (Fig.~\ref{fig:kTprime}) will both be reduced by energy loss in the medium to become $\hat{p}{'}_{Tt}$ and $\hat{p}{'}_{Ta}$ which will be measured by the di-hadron correlations with $p_{Tt}$ and $p_{Ta}$ in Au$+$Au collisions. The azimuthal angle between the di-jets, determined by the $\mean{k_{T}^2}$ in the original collision, should not change as both jets lose energy unless the medium induces multiple scattering from $\hat{q}$. Thus, without $\hat{q}$ and assuming the same fragmentation transverse momentum $\rms{j_T}$ in the original jets and those that have lost energy, the $p_{\rm out}$ between the away hadron with $p_{Ta}$ and the trigger hadron with  $p_{Tt}$ will not change;  but the $\mean{k{'}_{T}^2}$ will be reduced because the ratio of the away to the trigger jets $\hat{x}{'}_h=\hat{p}{'}_{Ta}/\hat{p}{'}_{Tt}$ will be reduced. Thus the calculation of $k{'}_T$ from the di-hadron p$+$p measurement to compare with Au$+$Au measurements with the same di-hadron $p_{Tt}$ and $p_{Ta}$ must use the values of $\hat{x}_h$, and $\mean{z_t}$ from the Au$+$Au measurement to compensate for the energy lost by the original dijet in p$+$p collisions.

The same values of $\hat{x}_h$, and $\mean{z_t}$ in Au$+$Au and p$+$p simplify Eqs.~\ref{eq:broadening} and \ref{eq:kTcalc} to:
\begin{equation}\mean{\hat{q} L}/2=\left[\frac{\hat{x}_h}{\mean{z_t}}\right]^2_{AA} \;\left[\frac{\mean{p^2_{\rm out}}_{AA} - \mean{p^2_{\rm out}}_{pp}}{x_h^2}\right] \label{eq:coolqhat}\end{equation}
\noindent
from which one could immediately get a reasonable answer for $\mean{\hat{q} L}/2$ from the $\mean{p^2_{\rm out}}$ results indicated on Fig.~\ref{fig:poutfitstostar} if the values of $\hat{x}_h$ and $\mean{z_t}$ in the Au$+$Au measurement are known.

\section{How to calculate $\mean{\hat{q} L}$ from the Au$+$Au (and p$+$p) measurements for di-hadrons with a trigger $p_{Tt}$ and away-side $p_{Ta}$ distribution.}
From Eq.~\ref{eq:coolqhat}, we need $\mean{p^2_{\rm out}}_{pp}$, $\mean{p^2_{\rm out}}_{AA}$ plus  $\hat{x}_h$ and $\mean{z_t}$ in Au$+$Au. This will be illustrated with the STAR data~\cite{STARPLB760}. 
\begin{enumerate}
\item[a)]{\bf $\hat{x}_h$ for a given $p_{Tt}$ can be calculated from the $p_{Ta}$ distribution:} 
The ratio of the away jet to the trigger jet transverse momenta, $\hat{x}_h=\hat{p}_{Tt}/\hat{p}_{Ta}$, can be calculated (Fig.~\ref{fig:xEfit}) from the  away particle $x_h=p_{Ta}/p_{Tt}$ distributions, which were also given in the STAR paper. The formula is~\cite{ppg029}, where $n$ is the power of the $p_T$ spectra:
    \begin{equation}
\left.{dP \over dp_{Ta}}\right|_{p_{Tt}}  = {N\,(n-1)}{1\over\hat{x}_h} {1\over {(1+ {x_h \over{\hat{x}_h}})^{n}}} \,  
\qquad . \label{eq:jetratio} \end{equation}
\begin{figure}[!h]  
\begin{center}
\raisebox{+0.2pc}{{\footnotesize a)}\hspace*{-0.05pc}\includegraphics[width=0.48\textwidth]{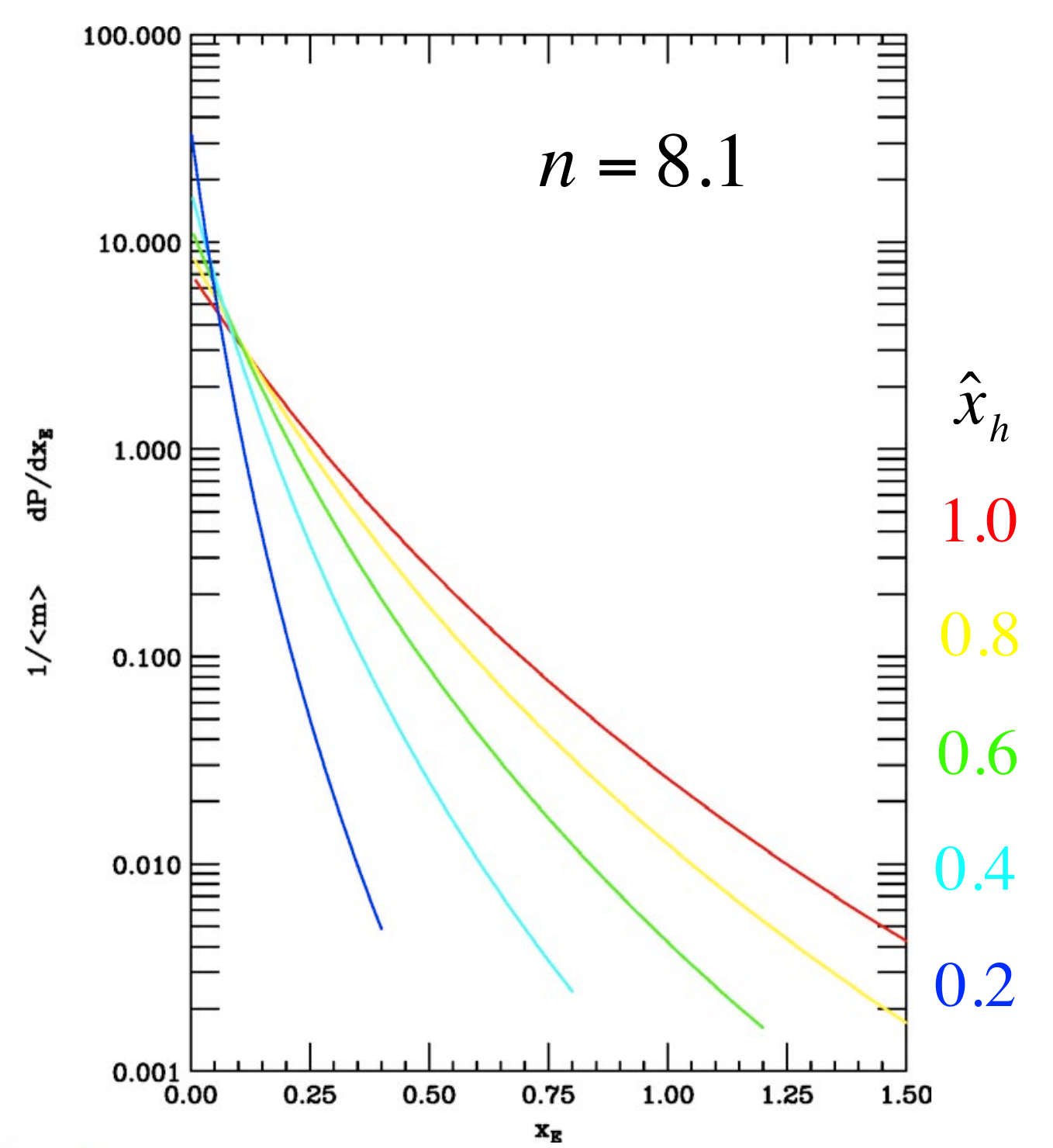}}
\raisebox{-0.0pc}{{\footnotesize b)}\hspace*{-0.05pc}\includegraphics[width=0.46\textwidth]{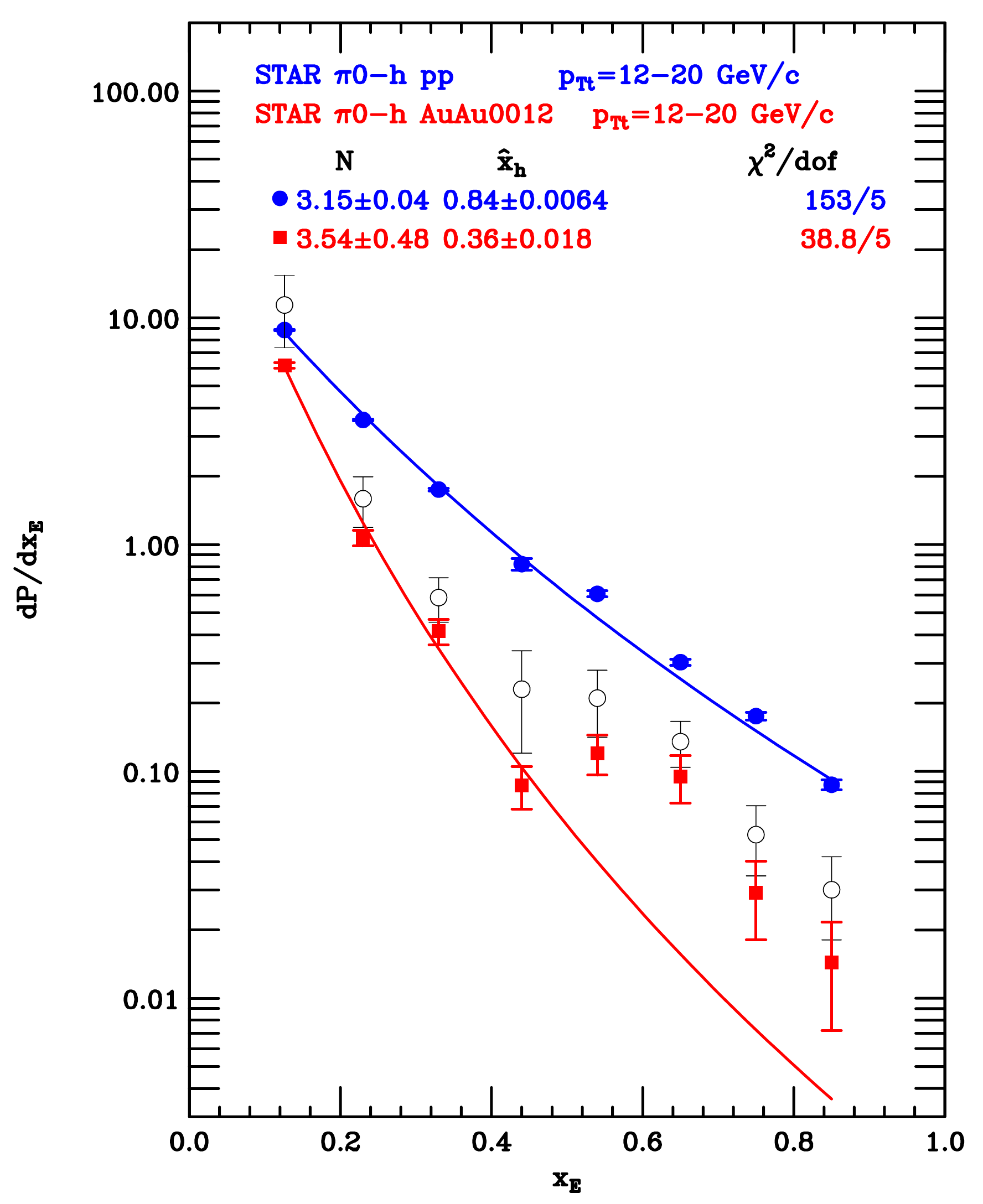}}
\caption[]{\footnotesize a) Plots of Eq.~\ref{eq:jetratio} for the values of $\hat{x}_h$ indicated.  b) Fits of Eq.~\ref{eq:jetratio}~\cite{MJTPLB771} to the STAR away-side $z_T$ distributions~\cite{STARPLB760} in  Au$+$Au 0-12\% centrality, and p$+$p, for $12<p_{Tt}<20$ GeV/c. The Au$+$Au curve is a fit with $\hat{x}^{AA}_h=0.36\pm 0.05$ with error corrected by $\sqrt{\chi^2/\rm{dof}}$. The points with the open circles are the $y_i$ and systematic errors $\sigma_{b_i}$ of the data points while the filled points are $y_i+\epsilon_b\sigma_{b_i}$ with errors $\tilde{\sigma}_i$ and $\epsilon_b=-1.3\pm0.5$.~\cite{MJTPLB771} }   
\label{fig:xEfit}\vspace*{-0.12in}
\end{center}
\end{figure} 
\item[b)]{\bf Fit the away-side peaks in the Au$+$Au and p$+$p correlation functions to gaussians in $\mathbf{p_{\rm out}}$:} 
The gaussian fit directly gives $\mean{p^2_{\rm out}}$ as was nicely shown for the STAR data in Fig.~\ref{fig:poutfitstostar}.
\item[c)]{\bf The power of hard scattering: the Bjorken parent-child relation and "trigger bias":}
The hard-scattering $p_T$ spectra, $d\sigma/p_T dp_T$, at RHIC in the range $3\leq p_T\lsim 20$ GeV/c for p$+$p and Au$+$Au for all centralities follow the same power law $1/p_T^n$ with $n=8.10\pm 0.05$~\cite{PXppg080}. This is why $R_{AA}(p_T)$ for $\pi^0$ and $\eta$ in Fig.~\ref{fig:suppression}b are relatively constant over the same $p_T$ range. 
The Bjorken parent-child relation~\cite{JacobPLC48} proved that the power $n$ in $p_T^{-n}$ is the same in the jet and fragment ($\pi^0$) $p_T$ spectra. This is why $\pi^0$ can be used in place of the parent jet. However because the trigger $\pi^0$ spectrum for a given $p_{Tt}$ in Au$+$Au for 0--10\% centrality is shifted down by $\delta p_T/p_T^{pp}=20\%$ in $p_T$ compared to p$+$p~\cite{ppg133}, the $\mean{z_t}$ for A$+$A and compensated p$+$p should be calculated~\cite{ppg029} from the measured p$+$p $\pi^0$ $p_T$ spectrum at $p_{Tt}^{pp}/(1-\delta p_T/p_T^{pp})$. (For the present discussion, STAR measured $\mean{z_t}=0.80\pm 0.05$ from their p$+$p data~\cite{STARPLB760}.)
\end{enumerate} 

This method enabled me to calculate $\mean{\hat{q} L}$ from the $\mean{p^2_{\rm out}}$ values indicated on Fig.~\ref{fig:poutfitstostar}, now with sensible results (Table~\ref{tab:star-PLB760}). The results in the two $p_{Ta}$ bins are at the edge of agreement, different by 2.4$\sigma$; but both are $>2.6 \sigma$ from zero. These results leave several open issues as mentioned in the abstract.  
   \begin{table}[!h]\vspace*{-0.0pc} 
\begin{center}
\caption[]{Tabulations for $\hat{q}$~\cite{MJTPLB771} from STAR $\pi^0$-h~data~\cite{STARPLB760}}
{\begin{tabular}{cccccc}  
\hline
\hline
$\sqsn=200$GeV &$\mean{p_{Tt}}$&$\mean{p_{Ta}}$&$\sqrt{\mean{k_{T}^2}}_{AA}$& $\sqrt{\mean{k{'}_{T}^2}}_{pp}$ & $\mean{\hat{q} L}$\\
 \hline
Reaction & GeV/c & GeV/c &GeV/c & GeV/c &GeV$^2$ \\
 \hline
Au$+$Au 0-12\%&14.71&1.72&$2.28\pm0.35$&$1.01\pm 0.18$& $8.41\pm2.66$\\
\hline
Au$+$Au 0-12\%&14.71 &3.75&$1.42\pm0.22$ &$1.08\pm 0.18$&$1.71\pm 0.67$\\
\hline\\[-1.0pc] 
\hline
\end{tabular}} \label{tab:star-PLB760}
\end{center}\vspace*{-0.1pc}
\end{table}\vspace*{-2.0pc}
\section{Homework}
\begin{figure}[!h]  
\begin{center}
\raisebox{+0.0pc}{{\footnotesize a)}\hspace*{-1.0pc}\includegraphics[width=0.455\textwidth]{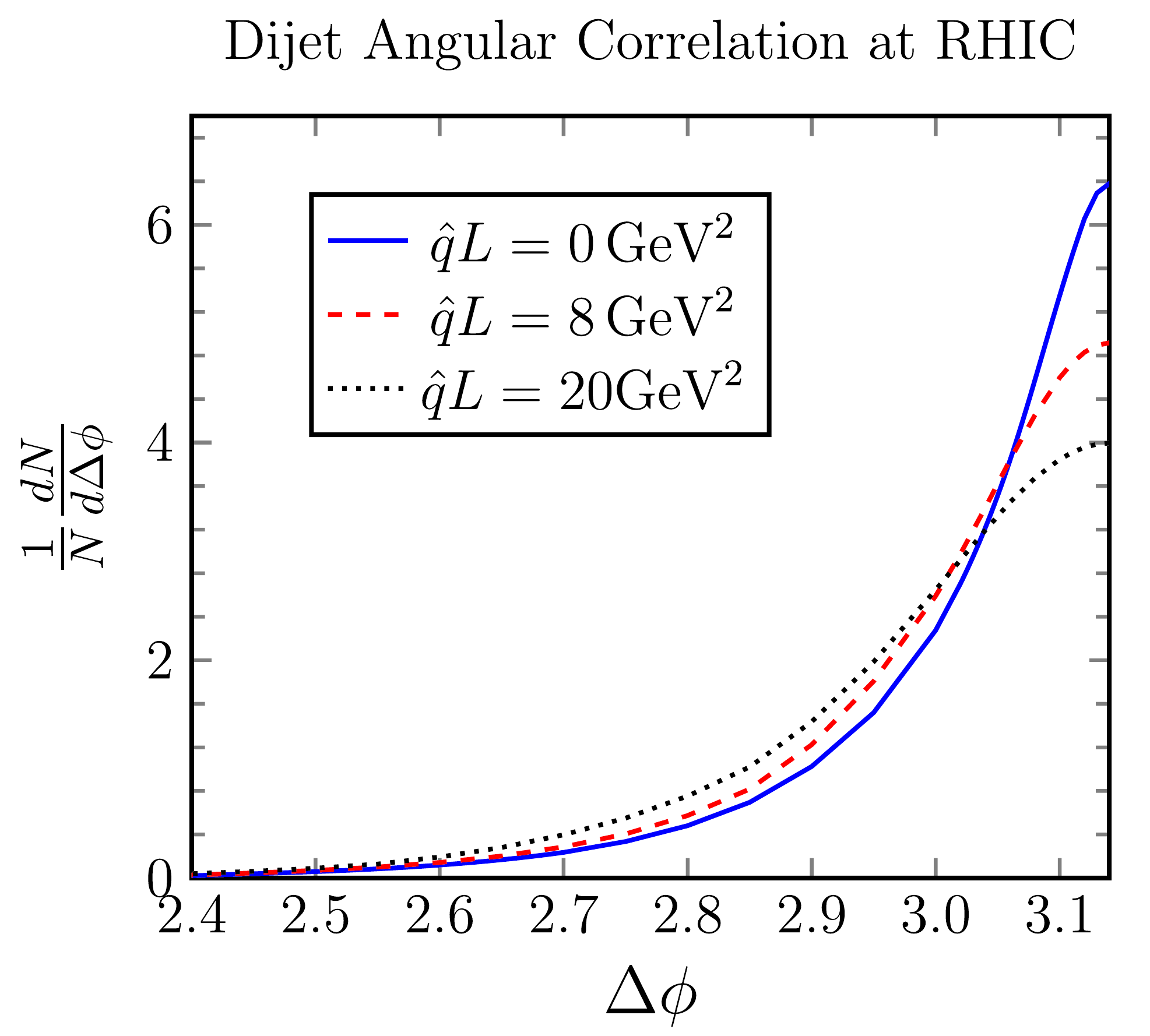}}\hspace*{1.0pc}
\raisebox{+0.0pc}{{\footnotesize b)}\hspace*{-1.0pc}\includegraphics[width=0.45\textwidth]{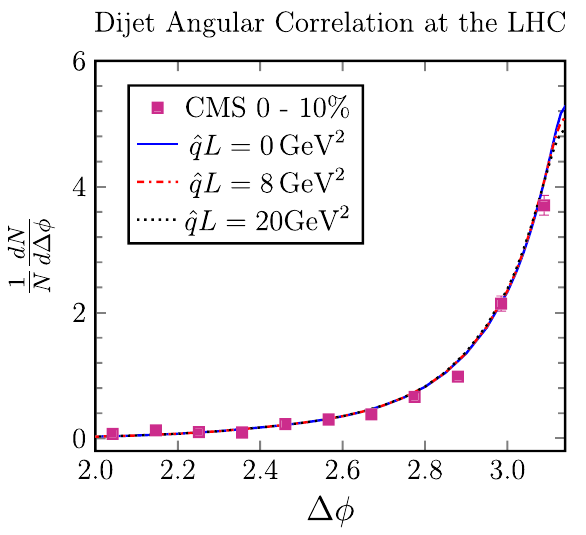}}
\caption[]{\footnotesize Prediction by Al Mueller and collaborators~\cite{MuellerPLB763} of the di-jet azimuthal decorrelation as a function of $\hat{q}L$ for a) 35 GeV jets at RHIC $\sqsn=200$ GeV; and b) 50 GeV jets at the LHC $\sqsn=2.76$ TeV where ``$p_T$ broadening effects are negligible"~\cite{MuellerPLB763}. }   
\label{fig:MuellerqhatL}
\end{center}
\end{figure}

However there is a nice prediction of $\Delta\phi$ for for 35 GeV Jets at RHIC~\cite{MuellerPLB763} for several values of $\mean{\hat{q} L}$ (Fig.~\ref{fig:MuellerqhatL}). An amusing test would be to see if the present method gives the same answers for $\mean{\hat{q} L}$ by calculating $\mean{p^2_{\rm out}}$ of the predictions.


\begin{thebibliography}{99}
\bibitem{BSZ2000} R.~Baier, D.~Schiff and B.~G.~Zakharov, {\it Energy Loss in Perturbative QCD},{\Journal{\ARNPS}{50}{37--69}{2000}} [{\tt hep-ph/0002198}].
\bibitem{MatsuiSatz} T.~Matsui and H.~Satz, {\it $J/\Psi$ Suppression by Quark-Gluon Plasma Formation},\Journal{\PLB}{178}{416--422}{1986}.
\bibitem{PXPRL88} K.~Adcox {\it et al.} (PHENIX), {\it Suppression of Hadrons with Large Transverse Momentum in Central Au + Au Collisions at $\sqsn=130$ GeV}, \Journal{\PRL}{88}{022301}{2002} [{\tt nucl-ex/0109003}]. 
\bibitem{MJTPLB771} M.~J.~Tannenbaum, {\it Measurement of $\hat{q}$ in Relativistic Heavy Ion Collisions using di-hadron correlations}, \Journal{\PLB}{771}{553--557}{2017} [{\tt arXiv:1702.00840}]
\bibitem{STARJhPRL112} L.~Adamczyk {\it et al.} (STAR), {\it Jet-Hadron Correlations in $\sqsn=200$ GeV p$+$p and Au$+$Au Collisions}, \Journal{\PRL}{112}{122301}{2014} [{\tt arXiv:1302.6184}] 
\bibitem{JacobsNPA956} P.~M.~Jacobs, A.~Schmah {\it et al.} (STAR), {\it Measurements of jet quenching with semi-inclusive charged jet distributions in Au+Au collisions at \sqsn=200 GeV},\Journal{\NPA}{956}{641--644}{2016} [{\tt arXiv:1512.08784}] 
\bibitem{FFFNPB128} R.~P.~Feynman, R.~D.~Field and G.~C.~Fox, {\it Correlations Among Particles and Jets Produced with Large Transverse Momenta},  \Journal{\NPB}{128}{1--65}{1977}. 
\bibitem{CCORjTkT} A.~L.~S.~Angelis, {\it et al.} (CCOR), {\it A Measurement of the Transverse Momenta of Partons, and of Jet Fragmentation as a Function of \sqs in pp Collisions}, \Journal{\PLB}{97}{163--168}{1980}.
\bibitem{ppg029} S.~S.~Adler, {\it et al.} (PHENIX), {\it Jet properties from dihadron correlations in p$+$p collisions at  \sqs=200 GeV}, {\Journal{\PRD}{74}{072002}{2006}}  [{\tt hep-ex/0605039}].
\bibitem{STARPLB760} L.~Adamczyk {\it et al.} (STAR), {\it Jet-like correlations with direct-photon and neutral-pion triggers at \sqsn=200 GeV}, \Journal{\PLB}{760}{689--696}{2016} [{\tt arXiv:1604.01117}]. 
\bibitem{PXppg080} A.~Adare, {\it et al.} (PHENIX). {\it Suppression Pattern of Neutral Pions at High Transverse Momentum in Au$+$Au Collisions at \sqsn=200 GeV and Constraints on Medium Transport Coefficients} \Journal{\PRL}{101}{232301}{2008} [{\tt arXiv:0801.4020}]
\bibitem{JacobPLC48} M.~Jacob and P.~V.~Landshoff, {\it Large Transverse Momentum and Jet Studies} \Journal{\PLC}{48}{285--350}{1978}.
\bibitem{ppg133} A.~Adare, {\it et al.} (PHENIX). {\it Neutral pion production with respect to centrality and reaction plane in Au + Au collisions at \sqsn=200 GeV}. \Journal{\PRC}{87}{034911}{2013} [{\tt arXiv:1208.2254}] 
\bibitem{MuellerPLB763} A.~H.~Mueller, {\it et al.} {\it Probing transverse momentum broadening in heavy ion collisions}. \Journal{\PLB}{763}{208--212}{2016}.  [{\tt arXiv:1604.04250}] 
\end{thebibliography}
\end{document}